\documentclass[12pt]{article}
\usepackage{sw20elba}
\usepackage{amsmath}
\usepackage{amsfonts}
\usepackage{amssymb}
\usepackage{graphicx}
\newtheorem{theorem}{Theorem}

\newtheorem{corollary}[theorem]{Corollary}

\newtheorem{lemma}[theorem]{Lemma}

\newtheorem{remark}[theorem]{Remark}

\begin{document}

\author{E. N. Dancer and S. P. Hastings\thanks{SPH, who was partially supported by the
National Science Foundation, wishes to thank the University of Sydney for its
hospitality during the time most of this work was done.}}
\title{On the Global Bifurcation Diagram for the One-Dimensional Ginzburg-Landau
Model of \\Superconductivity }
\date{February 15, 1999}
\maketitle
\begin{abstract}
Some new global results are given about solutions to the boundary value
problem for the Euler-Lagrange equations for the Ginzburg-Landau model of a
one-dimensional superconductor. The main advance is a proof that in some
parameter range there is a branch of asymmetric solutions connecting the
branch of symmetric solutions to the normal state. Also, simplified proofs are
presented for some local bifurcation results of Bolley and Helffer. These
proofs require no detailed asymptotics for the solutions of the linear
equations. Finally, an error in Odeh's work on this problem is discussed.
\end{abstract}

\subsection{Introduction}

In 1950 Ginzburg and Landau $\cite{gl}$ proposed a model for the
electromagnetic properties of a film of superconducting material of width $2d
$ subjected to a tangential external magnetic field. Under the assumption that
all quantities are functions only of the transverse coordinate, they proposed
that the electromagnetic properties of the superconducting material are
described by a pair $({\phi},{a})$ which minimizes the free energy functional
\[
G=\frac{1}{2d}\int_{-d}^{d}({\phi}^{2}({\phi}^{2}-2)+\frac{2({\phi}^{\prime
})^{2}}{\kappa^{2}}+2{\phi}^{2}{a}^{2}+2({a}^{\prime}-h)^{2})dx.
\]
The functional $G$ is now known as the Ginzburg-Landau energy and provides a
measure of the difference between normal and superconducting states of the
material. The variable ${\phi}$ is the ``order parameter'' which measures the
density of superconducting electrons, and ${a}$ is the magnetic field
potential. Also, $h$ is the external magnetic field, and $\kappa$ is the
dimensionless constant distinguishing different superconductors. So-called
``type I'' superconductors have $0<\kappa<\frac{1}{\sqrt{2}}$ while
$\kappa>\frac{1}{\sqrt{2}}$ for type II superconductors. (But this is really
only valid for large $d$; see \cite{afttr}.)

The existence of minimizers for the functional $G$ is proved in a standard
way, and such minimizers satisfy the following Euler-Lagrange boundary value
problem:
\begin{equation}
{\phi}^{\prime\prime}=\kappa^{2}{\phi}({\phi}^{2}+{a}^{2}-1)\label{eq1}%
\end{equation}%

\begin{equation}
{a}^{\prime\prime}={\phi}^{2}{a}\label{eq2}%
\end{equation}
with boundary conditions
\begin{equation}
{\phi}^{\prime}(\pm d)=0,\,\,\,{a}^{\prime}(\pm d)=h.\label{eq3}%
\end{equation}

It is not hard to show that the solutions of physical interest are such that
${\phi}>0$ on $[-d,d]$, and this is the only kind of solution we will consider
in this paper. Also, $h>0.$ Our goal is to determine for what values of
$h,\,\,d,$ and $\kappa$ the problem has solutions, and how many solutions
there are in various parameter ranges.

There are two kinds of solutions of interest, so-called symmetric solutions,
where ${\phi}(x)$ is an even function of $x$ while ${a}(x)$ is odd, and
asymmetric solutions, where these conditions are not satisfied. There is a
family of trivial solutions, called ``normal states'' of the form
\[
{\phi}(x)=0,\,\,\,\,{a}(x)=h(x+c),
\]
which are obviously symmetric when $c=0$ and asymmetric otherwise. In an early
paper $\cite{od}$ Odeh studied when non-trivial solutions may bifurcate off
these normal solutions. He concluded that symmetric solutions did bifurcate
from the branch of normal solutions, but as we shall see just before Lemma
\ref{lem2}, his argument had a flaw. He also considered whether asymmetric
solutions could bifurcate off the normal state, but reached no definitive conclusion.

Subsequently, Bolley and Helffer wrote a series of papers on the problem
\cite{bolley}, \cite{bh1}, \cite{bh2}, \cite{bh3}, \cite{bh4} and other
references cited in $\cite{bh3}$. They gave a quite thorough treatment of the
local bifurcations which can occur from the normal state, with these results
summarized in $\cite{bh3}$ and \cite{bh4}. Among many results, they gave the
correct formulation of when and how symmetric solutions bifurcate from the
normal state, and did not make the error made by Odeh, though they appear not
to have noticed the discrepancy with his assertions. However some of their
proofs are complicated, so we will give some simplifications. The proofs below
are self-contained, and in particular, we note that at least for the results
considered below, it is not necessary to use detailed asymptotics for the
parabolic cylinder functions which solve the relevant bifurcation equation.

A problem of particular physical interest is whether, as the strength of the
magnetic field is lowered, asymmetric solutions bifurcate from the normal
state before the symmetric solutions. This problem is discussed by Boeck and
Chapman \cite{chapman2} and by Aftalion and Troy \cite{afttr}. They relate
this question to the formation of vortices in the medium, a phenomenon that
can not be seen in the one-dimensional model. \ According to these authors, if
$d$ is neither too small nor too large, and if the asymmetric solutions
bifurcate first, then interference between the two symmetrically placed
solutions at either edge of the slab can produce a row of vortices down the
center of the slab. The only rigorous result on this problem is by Bolley and
Helffer, who show that when $d$ is sufficiently large, it is indeed the case
that $h_{a}>h_{s}$. This is one of the results for which we give a simpler
proof below. \ A related result in two dimensions with radial symmetry appears
in \cite{phillips}.

After formulating our results we received a new paper of Aftalion and Troy
\cite{afttr}, who did a thorough numerical study of how the bifurcation curves
change with $\kappa$ and $d.$ Based on these computations they make a variety
of conjectures, some of which are related to our work. \ They conjecture in
particular that $h_{a}>h_{s}$ for any $\left(  \kappa,d\right)  $ such that
asymmetric solutions exist. (This has not been proved, though it is well
accepted by physicists \cite{afttr}.)\ Subsequently Aftalion and Chapman have
used methods of matched asymptotic expansions to study some of the phenomena
found by Aftalion and Troy \cite{aftchap}\cite{aftchap2}. \ 

The first rigorous study of the global bifurcation diagram for symmetric
solutions was by Kwong $\cite{kw}.$ He proved that for any $\left(
\kappa,d\right)  $\ there is a unique curve of symmetric solutions, which can
be given in the form $h=h(\phi(0))$\ $\ $for $0<\phi(0)<1.$\ This curve is
smooth, and $h(1)=0,$\ $\ h(0)=h_{s}.$\ Hastings, Kwong and Troy studied the
nature of this curve for large $d,$\ showing that it has at least one minimum,
followed by at least one maximum, if $\kappa>\sqrt{1/2}$. \ This implies that
for some values of $h$\ there will be at least three solutions of the boundary
value problem in this range of $\kappa$ and $d.$\ They also showed that for
any fixed $\kappa\in(0,\frac{1}{\sqrt{2}}),$ if $d$ is sufficiently large then
for some range of $h$ there will be at least two solutions. \ \ More recently,
Aftalion and Troy proved that for sufficiently small $\kappa d,$ there is only
one symmetric solution, and there are no asymmetric solutions \cite{afttr2}.
\ (Numerically it appears that asymmetric solutions begin when $\kappa d$
reaches approximately .905 $\cite{chapman2}\cite{afttr}.)$%

%TCIMACRO{\FRAME{ftbpFU}{3.3434in}{2.8193in}{0pt}{\Qcb{The horizontal axis is
%$h$ and the vertical axis is $a(d).$ The solid curve is the branch of
%symmetric solutions while the dotted curve is the branch of asymmetric
%solutions. The end points of these curves, other than $(0,0),$ are
%bifurcations from a normal state. }}{}{dancer1.ps}%
%width 3.3434in;  height 2.8193in;  depth 0pt;  original-width 8.8194in;
%original-height 7.4261in;  cropleft "0";  croptop "1";  cropright "1";
%cropbottom "0";  filename 'dancer1.ps';file-properties "XNPEU";}}}%
%BeginExpansion
\begin{figure}
[ptb]
\begin{center}
\includegraphics[
height=2.8193in,
width=3.3434in
]%
{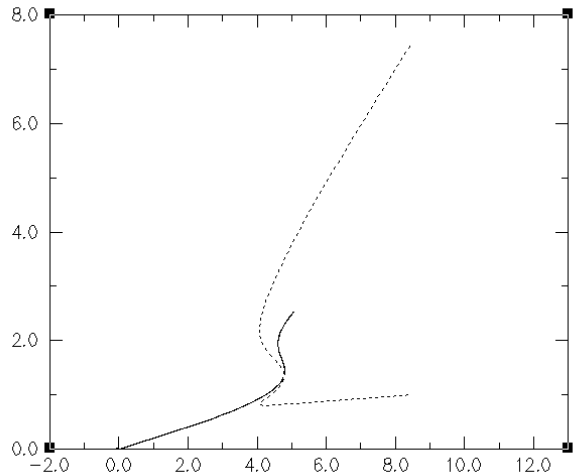}%
\caption{The horizontal axis is $h$ and the vertical axis is $a(d).$ The solid
curve is the branch of symmetric solutions while the dotted curve is the
branch of asymmetric solutions. The end points of these curves, other than
$(0,0),$ are bifurcations from a normal state. }%
\end{center}
\end{figure}
%EndExpansion
Up to now, very little has been done concerning the global structure of
asymmetric solutions (in the parameters $\kappa,d,h)$ or of bifurcations away
from the normal states. Some initial conjectures were made by Aftalion
$\cite{aft}.$ However a numerical study by Seydel $\cite{sy}$ shows that the
picture can be quite complicated. He considers only a single configuration,
namely $d=2.5,$ $\kappa=1,$ and presents essentially the graph in Figure 1, in
which $h$ is plotted against the value of ${a}$ at the right-hand end of the
interval $[-d,d].$ \ (Seydel uses $a(-d)$ instead of $a(d).$ There are a
number of possible ``bifurcation curves'' which one can draw for this problem.
For example, we could plot $h$ vs $\phi(0),$ as was done for symmetric
solutions in \cite{hkt}. \ We elect here to follow Seydel and plot $a(d)$ vs
$h.$ Either kind of curve gives the important information of how many
solutions there are for a given $h.)$

Among the features we see here are the existence of up to seven solutions for
a given $h,$ and bifurcation of asymmetric solutions from the symmetric
branch. It must be remembered, though, that asymmetric solutions occur in
pairs, and modulo a symmetric reflection, Seydel finds up to two asymmetric
solutions and three symmetric solutions for fixed $h.$

There are two obvious questions to ask concerning the Seydel result. How does
the picture change as $d$ and $\kappa$ vary, and what are the stability and
minimization properties of these solutions?

As stated above, the first of these questions was studied thoroughly in
\cite{afttr}. From their bifurcation diagrams, and results in \cite{bh4} one
can infer results about local stability near bifurcation points. Global
minimization was studied by Hastings and Troy $\cite{ht}.$ In addition to
demonstrating the existence of asymmetric solutions for large $d,$ they showed
that in some parameter range there are asymmetric solutions but no non-trivial
symmetric solutions. They also showed that the energy of the asymmetric
solutions can be negative, so that a global minimizer of the Ginzburg-Landau
functional $G$ must be asymmetric.

We have done some numerical work to consider the robustness of an asymmetric
global minimizer as we move into the region where both symmetric and
asymmetric solutions exist. \ More precisely, choosing the ``Seydel'' values
$\kappa=1,d=2.5,$\ so that asymmetric solutions exist, we started with
$h=h_{a},$\ the asymmetric bifurcation point. \ As $h$ is lowered, initially
there are only asymmetric solutions, and at points along the branch of
asymmetric solutions we evaluated the Ginzburg-Landau functional $G$\ and
found it to be negative, so that the solutions must be global minimizers.
\ (This is in accord with the result of Hastings and Troy.) \ Lowering
$h$\ further we reached the region where there are both symmetric and
asymmetric solutions. \ At a decreasing sequence of $h$\ values, we evaluated
$G$\ at each of the solutions existing for these values of $h;$\ up to
$5$\ distinct solutions. \ We found for a considerable distance down the
original curve of asymmetric solutions that $G$\ takes its minimum on this
curve. \ Thus, the minimization property of this asymmetric branch appears to
be very robust. \ 

However, this was a relatively crude examination, and by no means a thorough
study of the $\left(  \kappa,d\right)  $\ parameter space. \ In this paper we
consider only the existence of solutions of $\left(  \ref{eq1}\right)
-\left(  \ref{eq3}\right)  ,$ and not the stability properties of these solutions.

Turning to the structure of the bifurcation diagram as $\kappa$ and $d$
change, we have used the program $Auto$ $\cite{doe}$ to produce many
bifurcation diagrams for different parameter values. \ (Aftalion and Troy also
used Auto, independently.) \ Here are some samples:%

%TCIMACRO{\FRAME{ftbpFU}{3.4316in}{2.8193in}{0pt}{\Qcb{The conventions for
%Figure 1 are also used here. \ The pairs $\left(  \kappa,d\right)  $ are,
%clockwise from the top left: $\left(  .2,1\right)  ,\left(  .35,2\right)
%,\left(  1,1.25.\right)  ,\left(  .5,2\right)  ,\left(  .55,2\right)  ,\left(
%1,2.5\right)  .$}}{}{dancer2.ps}{\special{ language "Scientific Word";
%type "GRAPHIC";  maintain-aspect-ratio TRUE;  display "USEDEF";
%valid_file "F";  width 3.4316in;  height 2.8193in;  depth 0pt;
%original-width 9.6115in;  original-height 7.8845in;  cropleft "0";
%filename 'dancer2.ps';file-properties "XNPEU";}}}%
%BeginExpansion
\begin{figure}
[ptb]
\begin{center}
\includegraphics[
height=2.8193in,
width=3.4316in
]%
{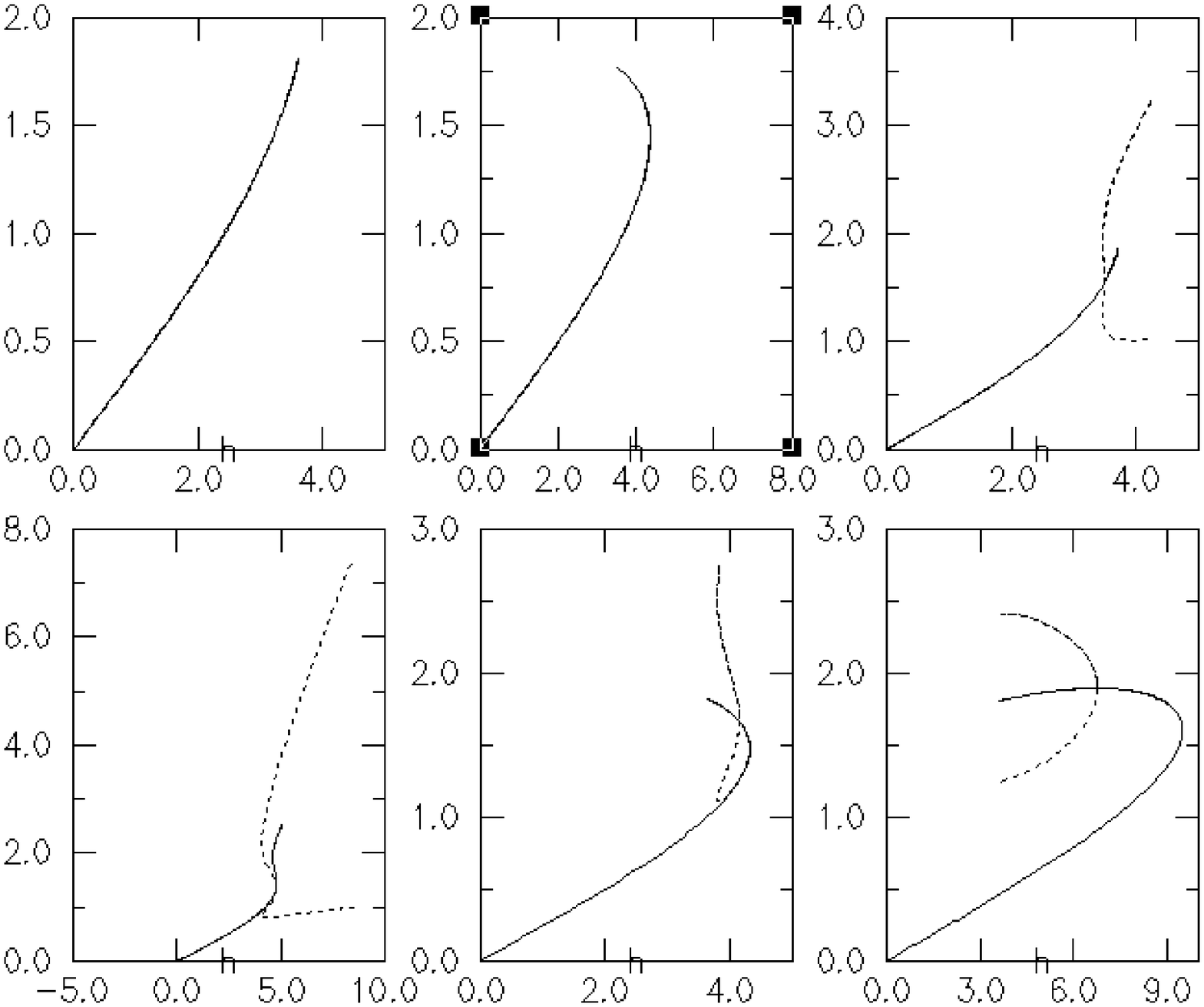}%
\caption{The conventions for Figure 1 are also used here. \ The pairs $\left(
\kappa,d\right)  $ are, clockwise from the top left: $\left(  .2,1\right)
,\left(  .35,2\right)  ,\left(  1,1.25.\right)  ,\left(  .5,2\right)  ,\left(
.55,2\right)  ,\left(  1,2.5\right)  .$}%
\end{center}
\end{figure}
%EndExpansion

\bigskip

From these pictures we see that there are at least the following possibilities:

\begin{enumerate}
\item  a single-valued curve of symmetric solutions in the $\left(
h,a(d)\right)  $ plane, and no asymmetric solutions,

\item  a single-valued curve of symmetric solutions, from which bifurcates a
C-shaped curve of asymmetric solutions,

\item  a C-shaped curve of symmetric solutions, no asymmetric solutions,

\item  a C-shaped curve of symmetric solutions and a C-shaped curve of
asymmetric solutions,

\item  a C-shaped curve of symmetric solutions, from which there bifurcates a
W-shaped curve of asymmetric solutions,

\item  an S-shaped curve of symmetric solutions from which there bifurcates a
W-shaped curve of asymmetric solutions.
\end{enumerate}

We also see that the asymmetric curves can bifurcate from various parts of the
C- or S-shaped symmetric curves. Some of these features, and others, are
discussed in more detail in \cite{afttr}.

\subsection{Statement of Results}

First we consider bifurcations from the normal state. Since the normal state
has ${\phi}=0,$ we rescale by letting
\[
\phi=\alpha\psi,
\]
where $\alpha=\phi(-d)$ and $\psi(-d)=1.$ Then,
\begin{equation}
\psi^{\prime\prime}=\kappa^{2}(\alpha^{2}\psi^{2}+a^{2}-1)\label{eq4}%
\end{equation}%

\begin{equation}
a^{\prime\prime}=\alpha^{2}\psi^{2}a\label{eq5}%
\end{equation}%

\begin{equation}
\psi(-d)=1,\psi^{\prime}(\pm d)=0,\,\,\,a^{\prime}(\pm d)=h.\label{eq6}%
\end{equation}
For $\alpha=0$ there is a family of solutions $(\psi_{0},h(x+c))$ where
$h=h(c)$ is chosen so that the linear problem
\begin{equation}
\psi_{0}^{\prime\prime}=\kappa^{2}(h^{2}(x+c)^{2}-1)\psi_{0},\,\,\,\label{eq7}%
\end{equation}%
\begin{equation}
\psi_{0}(-d)=1,\,\,\,\psi_{0}^{\prime}(\pm d)=0.\label{eq7a}%
\end{equation}
has a unique positive solution $\psi_{0}.$ From standard linear theory
$\cite{cl}$ we have

\begin{lemma}
\label{lem1}For each $c$ there is a unique $h=h(c)>0$ such that $\left(
\ref{eq7}\right)  -\left(  \ref{eq7a}\right)  $ has a positive solution. When
it exists, this solution is unique.
\end{lemma}

Thus $\left(  \ref{eq4}\right)  -\left(  \ref{eq6}\right)  $ is degenerate at
$\alpha=0,$ in the sense that there is a continuum of solutions, because $c$
is arbitrary.

\bigskip

To remove this degeneracy we reformulate the problem. \ Consider the equations
$\left(  \ref{eq4}\right)  -\left(  \ref{eq5}\right)  $ with initial
conditions $\psi(-d)=1,\psi^{\prime}(-d)=0,a\left(  -d\right)  =h(c-d),$ and
$a^{\prime}(-d)=h.$ Then consider $\psi^{\prime}(d)$ and $a^{\prime}(d)$ as
functions of $\left(  \alpha,c,h\right)  .$ We wish to solve the equations
\begin{equation}
\psi^{\prime}(d)=0,a^{\prime}(d)-h=0\label{revision7-1}%
\end{equation}
for $\left(  c,h\right)  $ as functions of $\alpha.$

Since
\[
a^{\prime}(d)-h=\int_{-d}^{d}\alpha^{2}\psi\left(  x\right)  ^{2}a(x)dx
\]
we replace $\left(  \ref{revision7-1}\right)  $ with
\begin{equation}
\psi^{\prime}(d)=0,\int_{-d}^{d}\psi\left(  x\right)  ^{2}%
a(x)dx=0\label{revision7-2}%
\end{equation}
We get a solution at $\alpha=0$ by setting $h=h(c)$ as given in Lemma 1 \ and
looking for values of $c$ \ such that
\begin{equation}
I(c):=\int_{-d}^{d}(x+c)\psi_{0}^{2}dx=0.\label{eq8}%
\end{equation}

Suppose that $\left(  \ref{eq8}\right)  $ is satisfied for some $c=c_{1}$.
\ This gives a solution to $\left(  \ref{revision7-2}\right)  $ for
$\alpha=0,$ and this solution can be extended to $\alpha>0$ provided that at
$\left(  \alpha,c,h\right)  =(0,c_{1},h(c_{1})$ the determinant
\[
\det\left(
\begin{array}
[c]{cc}%
\frac{\partial\psi_{0}^{\prime}(d)}{\partial h} & \frac{\partial\psi_{0}^{\prime}%
(d)}{\partial c}\\
\frac{\partial I}{\partial h} & I^{\prime}(c)
\end{array}
\right)
\]
is nonzero. \ We will see later (equation $\left(  \ref{14b}\right)  )$ that
$I(c)=0$ implies that $\frac{\partial\psi_{0}^{\prime}(d)}{\partial c}=0,$ and
standard theory implies that $\frac{\partial\psi_{0}^{\prime}(d)}{\partial h}%
\neq0.$ Hence, a unique branch of solutions bifurcates from $\left(
0,c_{1},h(c_{1}\right)  )$ provided that $I^{\prime}(c_{1})\neq0.$ \ 

For $c=0,$ $a=hx$ is an odd function, and this implies that $\psi_{0}$ is
even. Thus it is automatic that $I(0)=0.$ Further, \ for $\alpha>0$ we can
consider instead of $\left(  \ref{eq4}\right)  -\left(  \ref{eq6}\right)  $
the problem for symmetric solutions. \ This is $\left(  \ref{eq4}\right)
-\left(  \ref{eq5}\right)  $ on $\left[  0,d\right]  $ with $\psi^{\prime
}(0)=\psi^{\prime}(d)=0,$ $a(0)=0,$ $a^{\prime}(d)=h.$ Kwong's result in
$\cite{kw}$ shows that this has a unique solution for each $\alpha\in(0,1),$
defining $h$ as a function of $\alpha.$ If $I^{\prime}(0)\neq0,$ then this is
also the unique solution in a neighborhood of the bifurcation point $(0,c_1,h(c_1)$ to $\left(  \ref{eq4}\right)  -\left(  \ref{eq6}%
\right)  .$ Hence a unique solution, which is symmetric,
bifurcates from $\left(  0,0,h(0)\right)
$ \ when $I^{\prime}(0)\neq0.$ \ \ But this condition depends on $d.$ When it
is not satisfied we expect a more complicated picture.

In $\cite{od},$ Odeh claimed that $I^{\prime}(0)$ is always positive (for any
$d$). He may have thought this was so because he thought that $\frac
{\partial\psi_{0}}{\partial c}$ was an even function of $x$ when $c=0,$ but in
fact, this function is neither even nor odd. We have the following result,
also proved by Bolley$\cite{bh1}$, but with a much longer proof:

\begin{lemma}
\label{lem2} Suppose that for each positive $\kappa,$ and $d,$ and each $c,$
$h$ is chosen as in Lemma 1. Then for sufficiently small $\kappa d,$
$I^{\prime}(0)>0,$ while for sufficiently large $\kappa d,$ $I^{\prime}(0)<0.$
\end{lemma}

The (relatively short) proof will be given in section 3. As a consequence we have

\begin{theorem}
\label{th0} For any $(\kappa,d)$ there is a bifurcation of symmetric solutions
from the normal state. For sufficiently large $\kappa d$ and for sufficiently
small $\kappa d$ a unique curve of symmetric solutions bifurcates from the
normal state. In other words, for sufficiently small $\alpha,$ and for some
$\delta>0,\left(  \ref{eq4}\right)  -\left(  \ref{eq6}\right)  $ has a unique
solution with $|h-h(0)|<\delta,$ and this solution is symmetric.

Further, suppose that $(\kappa_{1},d_{1})$ and $(\kappa_{2},d_{2})$ are such
that $I^{\prime}(0)<0$ if $(\kappa,d)=(\kappa_{1},d_{1})$ and $I^{\prime
}(0)>0$ if $(\kappa,d)=(\kappa_{2},d_{2}),$ and assume that $(\kappa(t),d(t))$
is a real analytic curve $C$ joining $(\kappa_{1},d_{1})$ and $(\kappa
_{2},d_{2}),$ with $\kappa(0)=\kappa_{1},$ $d(0)=d_{1}$ and $\kappa
(1)=\kappa_{2}$ and $d(1)=d_{2}.$ Then there exists a $t_{0}\in(0,1) $ and
asymmetric solutions (which are nearly symmetric) arbitrarily close to
$(0,\tilde{h}_{0}x)$ with $h$ near $\tilde{h}_{0}$, $\kappa$ near
$\kappa(t_{0})$ and $d$ near $d(t_{0}).$ Here $\tilde{h}_{0}$ is the
eigenvalue found in Lemma \ref{lem1} for $(\kappa,d)=(\kappa(t_{0}),d(t_{0}))$
and $c=0.$
\end{theorem}

\begin{remark}
Note that, unlike \cite{bh1}, we do not need a transversality assumption, and
with care we could avoid assuming that the curve is analytic. Note also that
Lemma \ref{lem2} ensures that suitable points $(\kappa_{1},d_{1})$ and
$(\kappa_{2},d_{2})$ exist.
\end{remark}

The asymmetric solutions obtained in Theorem 3 are, at least initially, nearly
symmetric, since they start from $c=0.$ A different sort of bifurcation of
asymmetric solutions was obtained by Bolley and Helffer $\cite{bh3}$, and
independently, with a different proof, by Hastings and Troy $\cite{ht}.$ In
this case we consider a fixed large $\kappa d,$ and vary $c,$ looking for
other values of $c$ where $I(c)=0.$ The symmetry in the problem means we only
have to consider positive $c.$ It is obvious that for $c\geq d, $ $I(c)>0,$ so
using Lemma \ref{lem2} we have:

\begin{corollary}
\label{cor1}For sufficiently large $\kappa d,$ there is at least one $c>0$
where $I(c)=0.$
\end{corollary}

In fact, there is only one such $c$ and asymmetric bifurcation occurs at this
point. Thus, at this positive $c$ where $I(c)=0,$ we have $h=h_{a}.$ The
uniqueness of $\ $this positive $c$ was initially shown by Bolley and
Helffer\cite{bh3}, but here we will give a simpler, self-contained, proof.

\begin{theorem}
\label{unique} For sufficiently large $\kappa d$ there is exactly one
$c=c_{1}>0$ such that $I(c_{1})=0.$ (By symmetry there is also one negative $c
$ with this property.) \ Further, $h(c_{1})>h(0).$
\end{theorem}

Finally, it is not hard to show that bifurcation does not occur for small
$\kappa d$ \cite{bh3}.

Now we turn to more global results. The goal now is to show that bifurcation
of asymmetric solutions can occur from the interior of the symmetric branch,
rather than just from normal states, and show that the resulting branch of
asymmetric solutions can be continued in the $(h,a(-d))$ plane ( $d$ large and
fixed) to the asymmetric bifurcation point which was found in Theorem
\ref{unique}. To state this result we must first recall a result of Kwong
$\cite{kw}.$ This result is about symmetric solutions, and concerns the global
bifurcation curve of symmetric solutions, for any fixed $d$ and $\kappa.$
Since we are considering only symmetric solutions, we consider $\left(
\ref{eq1}\right)  $ with the following \textbf{initial} conditions:
\begin{equation}
\phi(0)=\alpha,\,\,\phi^{\prime}(0)=0,\,\,a(0)=0,\,\,a^{\prime}(0)=\delta
,\label{eq9}%
\end{equation}
where $\alpha\in(0,1)$ and $\delta>0$ are to be chosen such that $\phi>0$ on
$[0,d]$ and $\phi^{\prime}(d)=0.$ By continuing the resulting solution with
$\phi$ even and $a$ odd to the entire interval $[-d,d]$ we get a symmetric
solution to $\left(  \ref{eq1}\right)  ,\left(  \ref{eq2}\right)  .$ Kwong's
result is:

\begin{lemma}
\label{lem4}For each $\alpha\in(0,1]$ there is a unique $\delta>0$ such that
the solution of $\left(  \ref{eq1},\ref{eq8}\right)  $ is positive and
satisfies $\phi^{\prime}(d)=0.$
\end{lemma}

Hence, for each $\alpha$ we obtain a unique $h=a^{\prime}(d)$ and a unique
$a(d).$ Plotting $a(d)$ vs $h$ gives the global bifurcation curve for
symmetric solutions in the form referred to above. An alternative form as used
in $\cite{hkt}$ is to plot $h$ vs $\alpha.$ The main new result of this paper is:

\begin{theorem}
\label{th4}If the product $\kappa d$ is sufficiently large, then bifurcation
of asymmetric solutions occurs somewhere along the curve of symmetric
solutions. For any fixed $\kappa,$ if $d$ is sufficiently large, there is a
continuum of asymmetric solutions which connects the curve of symmetric
solutions to an asymmetric normal state.
\end{theorem}

\begin{remark}
Here we are identifying points in the $(a(d),h)$\ bifurcation diagram
corresponding to asymmetric pairs of solutions. The asymmetric normal state
referred to in this theorem must be the one discussed in Corollary \ref{cor1}
and Theorem \ref{unique},\ since to within a reflection there is only one
asymmetric bifurcation point from the normal state. \ By real analyticity the
continuum in this theorem contains a curve which is parametrized (in some
sense) by $h.$
\end{remark}

\begin{remark}
Also, it is expected that for $\kappa>\frac{1}{\sqrt{2}},$ if $d$ is
sufficiently large then the curve of symmetric solutions is S-shaped, so that
for some values of $h$ there are three solutions. In $\cite{hkt}$ it was shown
that there are \textbf{at least} three solutions, which is consistent with
this conjecture. Theorem \ref{th4} shows that bifurcation must occur somewhere
along this curve, but we are not able to prove anything about the location of
the bifurcation point on this curve. Similarly, for $\kappa<\frac{1}{\sqrt{2}%
}$ there are at least two solutions for large $d$, and the bifurcation seems
to occur on either of the two branches.
\end{remark}

\begin{remark}
A problem which we have not been able to solve is to determine the direction
of the bifurcation. As a result, we have not been able to prove that there are
some values of the parameters $(d,\kappa,h)$ where there are five distinct
solutions, as was seen in Seydel's original numerical result. In \cite{bh3}
there is a discussion of the stability of solutions bifurcating from the
normal states. This involves consideration of the energy functional $G,$ and
we have not studied this topic here.
\end{remark}

\section{ Proofs}

\subsection{Local Results}

In this section, we will denote $\psi_{0}$ by $\psi,$ since we are only
considering the linear equations. \ Thus, $\psi$ is assumed to satisfy
$\left(  \ref{eq7}\right)  -\left(  \ref{eq7a}\right)  .$

\bigskip

Proof of Lemma \ref{lem2}\newline 

\bigskip

In $\left(  \ref{eq7}\right)  -\left(  \ref{eq7a}\right)  $ we can make the
change of variables $x\rightarrow\kappa x,$ which removes the term $\kappa
^{2}$ from the differential equation, while replacing $d$ with $D=\kappa d,$
and also introducing a new $h$ and $c.$ The latter changes are immaterial in
this and subsequent results about bifurcation from the normal state, so
without loss of generality we will simply assume in $\left(  \ref{eq7}\right)
-\left(  \ref{eq7a}\right)  $ that $\kappa=1,$ and to remind us of this,
replace $d$ with $D.$ Multiply $\left(  \ref{eq7}\right)  $ by $\psi^{\prime}
$ and integrate by parts to get
\begin{equation}
2h^{2}I(c)=\left(  h^{2}(D+c)^{2}-1\right)  \psi(D)^{2}-\left(  h^{2}%
(-D+c)^{2}-1\right)  \psi(-D)^{2}.\label{eq10}%
\end{equation}

Consider $\left(  \ref{eq7}\right)  $ with the initial conditions
$\psi(-D)=1,\psi^{\prime}(-D)=0$ and denote the solution by $\psi(x,h,c).$ Let
$p=\frac{\partial\psi}{\partial h}$ and $q=\frac{\partial\psi}{\partial c}.$
Then
\begin{equation}
p^{\prime\prime}=\left(  h^{2}(x+c)^{2}-1\right)  p+2h(x+c)^{2}\psi
,\,\,\,\,p(-D)=p^{\prime}(-D)=0,\label{eq11}%
\end{equation}
and
\begin{equation}
q^{\prime\prime}=\left(  h^{2}(x+c)^{2}-1\right)  q+2h^{2}(x+c)\psi
,\,\,\,\,\,q(-D)=q^{\prime}(-D)=0.\label{eq12}%
\end{equation}

Lemma \ref{lem1} tells us that $\left(  \ref{eq7}\right)  -\left(
\ref{eq7a}\right)  $ define $h$ as a function of $c.$ We can also see this
locally by applying the implicit function theorem, solving the equation
\[
\psi^{\prime}(D,h,c)=0
\]
for $h$ as a function of $c.$ We can do this if $p^{\prime}(D)\neq0.$ Multiply
$\left(  \ref{eq11}\right)  $ by$\ \ \psi$ and $\left(  \ref{eq7}\right)  $ by
$p$ and integrate from $-D$ to $D.$ With the boundary and initial conditions
for $\psi$ and $q,$ this gives
\begin{equation}
\psi(D)p^{\prime}(D)=2h\int_{-D}^{D}(x+c)^{2}\psi(x)dx>0.\label{12aa}%
\end{equation}

Hence $h$ is a smooth function of $c.$ Further,
\begin{equation}
\frac{dh}{dc}=-\frac{q^{\prime}(D)}{p^{\prime}(D)}.\label{eq14aa}%
\end{equation}
Now multiply $\left(  \ref{eq12}\right)  $ by $\psi,$ $\left(  \ref{eq7}%
\right)  $ by $q,$ subtract and integrate, and use the boundary conditions
again to get
\begin{equation}
\psi(D)q^{\prime}(D)=2h^{2}\int_{-D}^{D}(x+c)\psi(x)^{2}dx=0.\label{14b}%
\end{equation}
This shows that $\frac{dh}{dc}=0$ whenever $I=0.$ In particular, this is true
at $c=0.$

Now differentiate $\left(  \ref{eq10}\right)  $ with respect to $c$ and then
set $c=0.$ Since, then, $\psi(\pm D)=1$, we get:
\begin{equation}
h^{2}I^{\prime}(0)=2h^{2}D+(h^{2}D^{2}-1)q(D).\label{eq13}%
\end{equation}

From $\left(  \ref{eq7}\right)  $ and $\left(  \ref{eq11}\right)  $ the
equation obtained at $c=0$ is
\begin{equation}
(\frac{q}{\psi})^{\prime}=\frac{\int_{-D}^{x}2h^{2}s\psi(s)^{2}ds}{\psi
(x)^{2}}.\label{eq14}%
\end{equation}
Since $\psi$ is an even function at $c=0,$ the right side of $\left(
\ref{eq14}\right)  $ is zero when $x=D.$ Further, the integrand is negative
for $s<0$ and positive for $s>0,$ and this implies that the integral is
strictly negative for $-D<x<D.$

Now return to $\left(  \ref{eq7}\right)  .$ We see that $\psi^{\prime\prime
}\geq-\psi,$ and since $\psi^{\prime}(0)=0,$ this leads to $\psi(x)\geq
\psi(0)\cos x$ on $\left[  0,D\right]  $ for small $D.$ It is therefore seen
that as $D\rightarrow0,$ $\max_{x\in\left[  0,D\right]  }\left|
\psi(x)-1\right|  \rightarrow0.$ Integrating the right side of the
differential equation for $\psi$ then shows that $\lim_{D\rightarrow0}%
hD=\sqrt{3},$ so $h\rightarrow\infty.$ From $\left(  \ref{eq14}\right)  $ it
follows that $\left(  \frac{q}{\psi}\right)  ^{\prime}=O(h)$ \ as
$D\rightarrow0,$ so $q(D)=O(hD)=O(1).$ Hence from $\left(  \ref{eq13}\right)
$ we see that for sufficiently small $D,$ $I^{\prime}(0)>0.$ \ 

To complete the proof of Lemma \ref{lem2} there remains to show that
$I^{\prime}(0)<0$ when $D$ is sufficiently large. We first need to show that
$h$ is bounded as $D\rightarrow\infty.$ In fact, it is well known, e.g.
\cite{ht}, that $h\rightarrow1$ as $D\rightarrow\infty,$ but the following
technique quickly shows that $h$ is at least bounded: Let $\rho=\frac
{\psi^{\prime}}{\psi}.$ Then
\begin{equation}
\rho^{\prime}=h^{2}x^{2}-1-\rho^{2},\rho(0)=\rho(D)=0,\label{eq14a}%
\end{equation}
with $\rho<0$ in $(0,D).$ (Remember that we are only considering $c=0$ here.)
It is easy to see from $\left(  \ref{eq14a}\right)  $ that if $h\rightarrow
\infty,$ then $D\rightarrow0.$ Hence, as $D\rightarrow\infty,$ $h$ must remain bounded.

It is also well known that $h>1,$ but for completeness here is a quick proof:
\ Compare $\rho$ from $\left(  \ref{eq14a}\right)  $ with the solution
$\sigma=-x$ of
\[
\sigma^{\prime}=x^{2}-1-\sigma^{2},\text{ \ }\sigma(0)=0.
\]
An easy comparison shows that if $h\leq1$ then $\rho(0)=\sigma(0)=0$ implies
that $\rho\leq\sigma$ for all $x\geq0,$ which contradicts $\rho(D)=0.$

Lemma \ref{lem2} now follows from $\left(  \ref{eq13}\right)  $ if we can show
that $q(D)$ does not tend to zero as $D$ tends to infinity. To do this we
again use $\left(  \ref{eq14}\right)  ,$ and also $\left(  \ref{eq7}\right)
$. Multiply $\left(  \ref{eq7}\right)  $ by $\psi^{\prime}$ and integrate from
$-D$ to $x.$ This, with $\left(  \ref{eq7a}\right)  $ and $\left(
\ref{eq14}\right)  ,$ leads to
\[
\left(  \frac{q}{\psi}\right)  ^{\prime}\leq(h^{2}x^{2}-1).
\]
Since $h$ is bounded for large $D,$ there is some interval around $x=0$ of
fixed length $\mu>0$ in which $\left(  \frac{q}{\psi}\right)  ^{\prime}%
\leq-\frac{1}{2}.$ In addition, $\frac{q}{\psi}$ is decreasing on the entire
interval $[-D,D]$, so $\frac{q(D)}{\psi(D)}\leq-\frac{1}{2}\mu.$ But
$\psi(D)=1$ when $c=0.$ Then $\left(  \ref{eq13}\right)  $ shows that
$I^{\prime}(0)<0$ for large $D.$ This proves Lemma \ref{lem2}.\newline 

Proof of Theorem \ref{th0}\newline 

Note that $I^{\prime}(0)$ is a function of $\kappa$ and $d.$ (We are no longer
assuming that $\kappa=1.)$ In fact, it is a real analytic function of $\kappa$
and $d,$ which is seen by observing that one can use the implicit function
theorem to prove that $\psi$ is a real analytic function of $(\kappa,d).$
\cite{cr} Thus $I^{\prime}(0)$ is a real analytic function of $t$ along the
curve $C$, since $C$ is real analytic. Since $I^{\prime}(0)$ does not vanish
identically, being nonzero at the endpoints of $C,$ its zeros are isolated.
Therefore there is a $t_{0}\in(0,1)$ such that $I^{\prime}(0)$ has a strict
change of sign as we cross $(\kappa(t_{0}),d(t_{0}))$ on $C.$

We now look for solutions $\phi=\alpha(\psi_{0}+w),$ $a=h(x+c)+\rho$ , where
$\psi_{0}$ is the eigenfunction at $t=t_{0},$ $h$ is close to $h_{0},$ and the
numbers $\alpha$ and $c$ and function $w$ and $\rho$ are small, and where $w$
is orthogonal to $\psi_{0}$ and $\rho$ is orthogonal to $1$, over $(-d,d).$
Equation $\left(  \ref{eq1}\right)  $ can be written as
\begin{equation}
-(\psi_{0}+w)^{\prime\prime}=\kappa^{2}(\psi_{0}+w)(1-a^{2}-\alpha^{2}%
(\psi_{0}+w)^{2}).\label{pr1}%
\end{equation}

We now use Fredholm alternative theory and the implicit function theorem in a
standard way (\cite{stak}, chapter 9) to show that for every $\left(
\alpha,c,h\right)  $ $\ $with $\alpha,c,$ and $(h-h_{0})$ sufficiently small,
there is a unique pair $(\beta,\gamma)$ such that the equations
\begin{align}
-(\psi_{0}+w)^{\prime\prime}  & =\kappa^{2}(\psi_{0}+w)(1-(h(x+c)+\rho
)^{2}-\alpha^{2}(\psi_{0}+w)^{2})+\beta\psi_{0}\label{pr2}\\
\rho^{\prime\prime}  & =\alpha^{2}(\psi_{0}+w)^{2}(h(x+c)+\rho)+\gamma
\label{pr3}%
\end{align}
have a unique solution $(w,\rho)$ such that $w^{\prime}(\pm d)=\rho^{\prime
}(\pm d)=0.$ In other words, we subtract off suitable multiples $\beta\psi
_{0}$ of $\psi_{0}$ and $\gamma\cdot1$ of $1$\ in equations $\left(
\ref{pr2}\right)  -\left(  \ref{pr3}\right)  $ respectively in order to find
solutions $(w,\rho)$ in the space
\begin{align}
\{w\in C^{1}[-d,d]:  & w\text{ is orthogonal to }\psi_{0},w^{\prime}(\pm
d)=0\}\times\{\rho\in C^{1}[-d,d]\\
& \text{and }\rho\text{ is orthogonal to }1,\rho^{\prime}(\pm d)=0\}\nonumber
\end{align}
We obtain $w,\rho,\beta,\gamma$ as smooth functions of $\alpha,c,h.$ Here we
are really solving a projected equation. \ This is a Lyapunov-Schmidt
reduction. \ We want to find solutions of $\left(  \ref{pr2}\right)  -\left(
\ref{pr3}\right)  $ with $\beta=\gamma=0.$

Let $\hat{\rho}$ denote the solution of
\begin{equation}
\hat{\rho}^{\prime\prime}=\psi_{0}^{2}x,\,\,\,\hat{\rho}^{\prime}(\pm
d)=0\label{pr3a}%
\end{equation}

with mean value zero. Using $\left(  \ref{eq2}\right)  $ it is easily shown
that
\[
\rho(\alpha,h,c)=h\alpha^{2}(\hat{\rho}+o(1))
\]
as $(\alpha,h,c)\rightarrow(0,h_{0},0).$ Note here that $s\psi_{0}^{2}(s)$ has
mean value zero. Hence,
\[
a^{2}=(h(x+c)+\alpha^{2}(\hat{\rho}+o(1)))^{2}=h^{2}(x+c)^{2}+2\alpha
^{2}h(x+c)\hat{\rho}+\text{terms of higher order.}
\]
(That is, higher order in $\alpha.$) \ Note that as $(\alpha,h,c)\rightarrow
(0,h_{0},0),$ $w$ and $\rho$ tend to zero uniformly in $[-d,d].$

Now multiply $\left(  \ref{pr2}\right)  $ by $\psi_{0}$ and integrate, using
the orthogonality of $w$ and $\psi_{0},$ to get
\begin{equation}
(h^{2}-h_{0}^{2})(\int_{-d}^{d}s^{2}\psi_{0}(s)^{2}ds+o(1))ds\ +\,\int
_{-d}^{d}o(1)ds=\beta\int_{-d}^{d}\psi_{0}(s)^{2}ds,\label{pr4}%
\end{equation}
where the $o(1)$ terms are terms in $w$ and $\rho$ which are smooth and tend
to zero as $(\alpha,h,c)\rightarrow(0,h_{0},0).$ (Also, the derivative with
respect to $h$ of the second $o(1)$ term is small if $\alpha$ is small, \ $h$
is near $h_{0}$ and $c$ is small.) To obtain a solution of $\left(
\ref{pr1}\right)  $ we set $\beta=0$ in $\left(  \ref{pr4}\right)  $ and use
the implicit function theorem to solve this equation for $h $ as a function of
$(\alpha,c)$ near $h=h_{0},\alpha=c=0$ . The coefficient of $(h^{2}-h_{0}%
^{2})$ in the first term on the left of $\left(  \ref{pr4}\right)  $is of
order 1. For each small $\alpha\neq0$ and each small $c$ there will be a
unique $h=h(\alpha,c)$ near to $h_{0}$ for which $\left(  \ref{pr4}\right)  $
implies that $\beta=0$.

Now integrate $\left(  \ref{eq2}\right)  $ over $[-d,d]$ and use the boundary
conditions to obtain, upon dividing by $\alpha^{2}h$ the equation
\[
R(\alpha,h,c)\equiv\frac{1}{\alpha^{2}h}\int_{-d}^{d}\phi^{2}a\,\,dx=\int
_{-d}^{d}((x+c)\psi_{0}^{2}(x)+o(1))dx=0
\]
where the small $o(1)$ term is a smooth function of $\alpha,h,c$ and is zero
if $\alpha=c=h-h_{0}=0.$ Note that $R(\alpha,h,c)=0$ is equivalent to
$\gamma=0.$ Also, $R(0,h_{0},c)=I(c).$ Letting $R_{3}^{\prime}(\alpha
,h,c)=\frac{\partial R}{\partial c},$ we have for large $\kappa d,$ by Lemma
\ref{lem2}, that $R_{3}^{\prime}(0,h_{0},0)<0,$ so $R_{3}^{\prime}%
(\alpha,h,c)<0$ if $\alpha$ and $c$ are small and $h$ is close to $h_{0}.$
Also, $R(\alpha,h,0)=0$ because then both $w$ and $\psi_{0}$ are even
functions of $x$ and $a$ is an odd function of $x.$ Hence, if $I^{\prime
}(0)<0,$ then $R(\alpha,h,c)<0$ if $\alpha$ is small, $c$ is positive and
small and $h$ is near to $h_{0}.$ Similarly, if $I^{\prime}(0)>0,$ then
$R(\alpha,h,c)>0$ if $\alpha$ is small, $c$ is small and positive, and $h$ is
near to $h_{0}.$

We now solve $\left(  \ref{pr4}\right)  $ for $h$ as a function of $c$ and
$\alpha,$ for small positive $\alpha$ and $c.$ (We do not know the sign of
$h-h_{0}.$) We do this for $(\kappa,d)$ on the curve $C$ with $(\kappa,d)$
close to $(\kappa(t_{0}),d(t_{0})).$ On this curve, on one side of
$(\kappa(t_{0}),d(t_{0})),$ $I^{\prime}(0)<0$ and hence $R(\alpha
,h(\alpha,c),c)<0 $ if $\alpha$ and $c$ are small and $c>0.$ How small
$\alpha$ and $c$ must be depends on the particular point on the curve $C.$ For
a given $t_{1}<t_{0} $ but close to $t_{0},$ find positive $\alpha_{1}$ and
$c_{1}$ such that $R(\alpha,h(\alpha,c),c)<0$ if $0<\alpha\leq\alpha_{1}$ and
$0<c\leq c_{1}.$ Choose a fixed $t_{2}>t_{0}$ but close to $t_{0},$ and then
lower $\alpha_{1}$ and $c_{1}$ if necessary so that $R(\alpha,h(\alpha
,c),c)>0$ if $0<\alpha\leq\alpha_{1}$ and $0<c\leq c_{1}.$Then somewhere
between $t_{1}$ and $t_{2}$, as we keep $\alpha$ and $c$ nonzero and fixed in
$(0,\alpha_{1}]$ and $(0,c_{1}]$ and move along $C,$ $R$ must equal 0, which
gives the required solution. (Note that $h_{0}$ varies continuously with
$(\kappa,d)$ but this does not affect the argument since everything varies
continuously along $C.$ ) Since we can choose $c$ arbitrarily small, the
solutions will be nearly symmetric.\linebreak 

Proof of Theorem \ref{unique}.\newline 

\bigskip

As previously, for the linearized problem $\left(  \ref{eq7}\right)  -\left(
\ref{eq8}\right)  $ we can rescale to eliminate $\kappa,$ so we will again
assume that $\kappa=1$ and replace $d$ with $D.$ We saw in the proof of Lemma
\ref{lem2} that for large $D,$ $I(0)=0,$ $I^{\prime}(0)<0,$ and that
consequently Corollary \ref{cor1} holds. Further, the definition of $I(c)$
shows that $I(c)>0$ for $c\geq D.$

From $\left(  \ref{eq7}\right)  -\left(  \ref{eq7a}\right)  ,$ with $\psi>0$,
we see that $\tau(x)=h(x+c)^{2}-1$ must change sign in $(-D,D),$ and from
$\left(  \ref{eq10}\right)  $ it follows that if $I(c)=0,$ then $\tau(-D)$ and
$\tau(D)$ must have the same sign, so $\tau()$ has exactly two zeros in
$(-D,D),$ with $\psi^{\prime\prime}(x)$ changing from positive to negative and
back to positive as $x$ increases from $-D$ to $D.$ Therefore, $\psi$ has a
local maximum at some $x_{0}\in(-D,D),$ with $\psi^{\prime}>0$ on $(-D,x_{0})$
and $\psi^{\prime}<0$ on $(x_{0,}D).$

\begin{lemma}
\label{lemunique1}For any $c\in(0,D),$ $h>1.$
\end{lemma}

Proof: Let $\rho(x)=\frac{\psi^{\prime}(x)}{\psi(x)}.$ Then $\rho
(-D)=\rho(x_{0})=\rho(D)=0.$ Also,
\begin{equation}
\rho^{\prime}=h^{2}(x+c)^{2}-1-\rho^{2}.\label{equn1}%
\end{equation}

Further, let $\sigma(x)=-x-c.$ Then
\begin{equation}
\sigma^{\prime}=(x+c)^{2}-1-\sigma^{2},\,\,\,\sigma(-c)=0.\label{equn2}%
\end{equation}

But $\rho(-D)<\sigma(-D),$ and an easy comparison of $\left(  \ref{equn1}%
\right)  $ and $\left(  \ref{equn2}\right)  $ shows that if $h\leq1,$ then
$\rho<\sigma$ on $[-D,D],$ so $\rho(D)<0$, a contradiction. This proves Lemma
\ref{lemunique1}.

\begin{lemma}
\label{lemunique2} For sufficiently large $D,$ $\left(  \ref{eq7}\right)
-\left(  \ref{eq8}\right)  $ has no solution with $\psi>0$ and $0<c\leq
\frac{D}{2}.$
\end{lemma}

Proof: Recall that $I(0)=0,$ $I^{\prime}(0)<0.$ Let $c_{1}=\inf
\{c>0|I(c)=0\}.$ Then $I^{\prime}(c_{1})\geq0.$ Assume that $c_{1}\leq\frac
{D}{2}.$ We will obtain a contradiction by differentiating $\left(
\ref{eq10}\right)  $ with respect to $c$ and setting $c=c_{1}.$

Recall that $\frac{dh}{dc}=0$ whenever $I(c)=0$. Therefore, from $\left(
\ref{eq10}\right)  $ we get
\begin{equation}%
\begin{array}
[c]{l}%
2h^{2}I^{\prime}(c_{1})=2h^{2}(D+c_{1})\psi(D)^{2}\\
\,\,\,\,\,\,\,\,\,\,\,\,\,\,\,\,\,\,\,\,\,\,\,\,+2(h^{2}(D+c_{1})^{2}%
-1)\psi(D)q(D)+2h^{2}(D-c_{1})\psi(-D)^{2}%
\end{array}
\label{equn3}%
\end{equation}
\bigskip where $q=\frac{\partial\psi}{\partial c}$, so $q$\ satisfies $\left(
\ref{eq12}\right)  .$\ From $\left(  \ref{eq12}\right)  $\ and $\left(
\ref{eq7}\right)  -\left(  \ref{eq7a}\right)  $\ we obtain $\left(
\ref{eq14}\right)  $\ with $s$\ replaced by $(s+c_{1}),$\ and as in the proof
of Lemma \ref{lem2}, we then see that $\left(  \frac{q}{\psi}\right)
^{\prime}<0$\ in $\left(  -D,D\right)  $\ and $\left(  \frac{q}{\psi}\right)
^{\prime}\leq h^{2}(x+c_{1})^{2}-1.$\ Also, as in Lemma \ref{lem2} it is seen
that $h$\ \ is bounded as $D\rightarrow\infty,$\ and this leads to a negative
upper bound of the form
\begin{equation}
\left(  \frac{q(D)}{\psi(D)}\right)  \leq-\eta<0,\label{equn6}%
\end{equation}
where $\eta$\ is independent of $D$\ and $c_{1}\in\left[  0,\frac{D}%
{2}\right]  .$

\bigskip\ From $I(c_{1})=0$\ and $\left(  \ref{eq10}\right)  $\ we obtain
\[
\psi(D)^{2}=\frac{h^{2}(D-c_{1})^{2}-1}{h^{2}(D+c_{1})^{2}-1}\psi(-D)^{2}%
\geq\frac{h^{2}(\frac{D}{2})^{2}-1}{2h^{2}D^{2}-1}\psi(-D)^{2}.
\]
Hence,\ $\frac{\psi(-D)^{2}}{\psi(D)^{2}}$\ is bounded as $D\rightarrow
\infty.$\ Then $\left(  \ref{equn6}\right)  $\ and $\left(  \ref{equn3}%
\right)  $ show that $I^{\prime}(c_{1})<0$\ for sufficiently large $D.$\ This
contradiction proves Lemma $\ref{lemunique2}$.

\bigskip

Continuing with the proof of Theorem \ref{unique}, we now assume that
$c\geq\frac{D}{2}.$ The result will follow if we can show that $I^{\prime
}(c_{1})>0$ for any solution of $\left(  \ref{eq7}\right)  -\left(
\ref{eq8}\right)  $ in this range of $c.$ As before, we assume that
$(c_{1},h,\psi)$ is a solution, and that $\psi$ has its maximum over $[-D,D]$
at $x_{0}\in(-D,D).$ Also, as before, we know that $|x_{0}+c_{1}|<1,$ so
$x_{0}\leq-\frac{D}{2}+1.$ We translate the origin to $x_{0},$ letting
$\psi(x)=\chi(x-x_{0})=\chi(y),$ so that
\[
\chi^{\prime\prime}(y)=(h^{2}(y+x_{0}+c_{1})^{2}-1)\chi(y).
\]

Now let $\rho(y)=\frac{\chi^{\prime}(y)}{\chi(y)}$ (a shift from the previous
$\rho)$ and let $\omega(y)=-\rho(-y).$ Then
\begin{equation}
\rho^{\prime}=h^{2}(y+x_{0}+c_{1})^{2}-1-\rho^{2},\,\,\,\rho(0)=\rho
(D-x_{0})=0\label{equn7}%
\end{equation}
and
\begin{equation}
\omega^{\prime}=h^{2}(y-x_{0}-c_{1})^{2}-1-\omega^{2},\,\,\,w(0)=w(D+x_{0}%
)=0.\label{equn8}%
\end{equation}
It is important to recall that $x_{0}<-\frac{D}{2}+1.$

We now need estimates on $\rho$ and $\omega,$ which we obtain from the
following result:

\begin{lemma}
\label{lemun2} Suppose that for some constants $\delta$ and $\Delta,$ with
$h|\delta|\leq1$ and $\Delta$ large, $\eta(\cdot)$ solves the boundary value
problem
\begin{equation}
\eta^{\prime}(y)=h^{2}(y+\delta)^{2}-1-\eta^{2},\,\,\eta(0)=\eta
(\Delta)=0,\,\,\eta<0\text{ in }(0,\Delta).\label{equn9}%
\end{equation}
Then,
\begin{equation}
\eta(y)>-hy-\beta\label{equn10}%
\end{equation}
on $[0,\Delta]$, where $\beta=\sqrt{h-1+h^{2}\delta^{2}},$ and
\begin{equation}
\eta(y)<-h(y+\delta)+1\label{equn11}%
\end{equation}
on $[0,\Delta-1].$
\end{lemma}

Proof:

Inequality $\left(  \ref{equn10}\right)  $ follows by assuming equality at
some $y_{1}\in(0,\Delta)$ and using $\left(  \ref{equn9}\right)  $ to show
that $\eta^{\prime}(y_{1})<-h.$ This would imply that $\eta(y)<-hy-\beta$ for
$y>y_{1,}$ so that $\eta$ could not vanish at $\Delta.$ Next, observe that
$\left(  \ref{equn11}\right)  $ holds as long as $h(y+\delta)\leq1,$ since
$\eta<0$ on $(0,\Delta).$ If equality holds at some $y_{1}>0,$ then
$h(y_{1}+\delta)>1$ and $\eta^{\prime}(y_{1})=-2+2h(y_{1}+\delta)>0.$ Also,
$\eta<0$ and $\eta^{\prime}>0$ imply that $\eta^{\prime\prime}>0. $ Hence,
$\eta^{\prime}>-2+2h(y_{1}+\delta)>0$ as long as $\eta<0,$ and if $\eta<0$ on
$\left[  y_{1},y_{1}+1\right]  ,$ then
\[
\eta(y_{1}+1)\geq\eta(y_{1})-2+2h(y_{1}+\delta)=-1+h(y_{1}+\delta)>0.
\]
This contradiction shows that $\eta=0$ somewhere in $\left[  y_{1}%
,y_{1}+1\right]  .$ If $y_{1}\leq\Delta-1,$ then $\eta=0$ before $y=\Delta,$ a
contradiction which proves Lemma \ref{lemun2}.

Applying $\left(  \ref{equn11}\right)  $ to $\rho$ with $\delta=x_{0}+c_{1}$
\ and $\Delta=D-x_{0}$ shows that
\[
\psi(D)=\chi(D-x_{0})\leq\psi(x_{0})e^{-\frac{h}{2}(D+c_{1}-1)^{2}+\frac{h}%
{2}(x_{0}+c_{1})^{2}+D-x_{0}-1},
\]
while applying $\left(  \ref{equn10}\right)  $ to $\omega$ with $\delta
=-x_{0}-c_{1}$ gives
\[
\psi(-D)\geq\psi(x_{0})e^{-\frac{h}{2}(D+x_{0})^{2}-\beta(D+x_{0})}.
\]
Combining these and noticing that $2Dx_{0}<-2x_{0}^{2}$ and $\beta<h,$ we find
that for $c_{1}\geq\frac{D}{2},$
\begin{equation}
\frac{\psi(D)}{\psi(-D)}\leq e^{-rhD^{2}}\label{equn13}%
\end{equation}
for some $r$ which is independent of $c_{1},$ $h,$ and $D.$

We now return to $\left(  \ref{equn3}\right)  .$ Since $\psi^{\prime\prime
}(-D)\geq0,$ we have $h^{2}(D-c_{1})\geq h.$ Further, the proof (following
equation $\left(  \ref{eq14}\right)  $ that when $c=0,$ $h$ is bounded as
$D\rightarrow\infty$ easily extends to $c\geq0,$ since $x_{0}\leq-c+1.$ From
$\left(  \ref{equn13}\right)  $ it follows that a bound of the form
\begin{equation}
q(D)\geq-Lh^{m}D^{n}\psi(D)\label{revised2}%
\end{equation}
for some $L>0$ independent of $c_{1},h,$ or $D$ will imply that $I^{\prime
}(c_{1})>0$ for large $D.$ There are two cases to consider, namely,
$-c_{1}-1\leq x_{0}\leq-c_{1}$ and $-c_{1}$ $\leq x_{0}\leq-c_{1}+1.$ We
consider the first, the two cases being similar. Repeating the derivation of
$\left(  \ref{eq14}\right)  $ \ for $c_{1}>0$ we obtain that
\begin{equation}
\frac{q(D)}{\psi(D)}=2h^{2}(\int_{-D}^{x_{0}}+\int_{x_{0}}^{-c_{1}}+\int
_{-c}^{D}\frac{1}{\psi(x)^{2}}\int_{-D}^{x}(s+c_{1})\psi(s)^{2}%
dsdx).\label{revised1}%
\end{equation}
In the first of the three integrals with respect to $x,$ $-D\leq s\leq x\leq
x_{0},$ so $\psi(s)\leq\psi(x),$ and this term contributes less than
$O(D^{3})$ to $\frac{q(D)}{\psi(D)}$ as $D\rightarrow\infty.$ In the third of
the three integrals, we use the fact that $I(c_{1})=0$ to write
\begin{equation}
\int_{-c_{1}}^{D}\frac{1}{\psi(x)^{2}}\int_{-D}^{x}(s+c_{1})\psi
(s)^{2}dsdx=-\int_{-c_{1}}^{D}\frac{1}{\psi(x)^{2}}\int_{x}^{D}(s+c_{1}%
)\psi(s)^{2}dsdx,\label{equn14}%
\end{equation}
and because $-c_{1}\geq x_{0}$ we again have $\psi(s)\leq\psi(x)$ and get a
contribution $O(D^{3}).$ The second term in $\left(  \ref{revised1}\right)  $
is bounded by
\[
\int_{x_{0}}^{x_{0}+1}\frac{\psi(x_{0})^{2}}{\psi(x_{0}+1)^{2}}\int_{-D}%
^{x}(s+c_{1})dsdx.
\]
Again we let $\rho=\frac{\psi^{\prime}(x-x_{0})}{\psi(x-x_{0})}$ and note that
$\rho^{\prime}\geq-1-\rho^{2},$ $\rho(0)=0.$ This gives a bound $\frac
{\psi(x_{0})}{\psi(x_{0}+1)}\leq e^{\delta}$ for some $\delta$ independent of
$D,$ $h,$ or $c_{1},$ and so the second term in $\left(  \ref{revised1}%
\right)  $ is $O(D^{2})$ as $D\rightarrow\infty.$

This proves the desired bound $\left(  \ref{revised2}\right)  .$ Hence
$\left(  \ref{equn13}\right)  $ shows that the dominant term in $\left(
\ref{equn3}\right)  $ is the last one, proving that $I^{\prime}(c_{1})>0$ for
large $D,$ if $D$ is sufficiently large and $I(c_{1})=0.$ Hence there is a
unique $c_{1}>0$ with $I(c_{1})=0.$ It follows from $\left(  \ref{12aa}%
\right)  -\left(  \ref{14b}\right)  $ that $h(c_{1})>h(0),$ since $I(c)<0$ for
$0<c<c_{1}.$ This completes the proof of Theorem \ref{unique}.

\subsection{Proofs of Global Results}

We now turn to the global results, about bifurcation from the curve of
symmetric solutions. We consider the full problem $\left(  \ref{eq1}\right)
-\left(  \ref{eq3}\right)  .$ The symmetric problem can be studied on the
interval $[0,d].$ Let $(\phi(x,\alpha,\delta),a(x,\alpha,\delta))$ denote the
solution of $\left(  \ref{eq1}\right)  -\left(  \ref{eq2}\right)  $ which
satisfies the initial conditions $\phi(0)=\alpha,$ $\phi^{\prime}(0)=0,$
$a(0)=0,$ $a^{\prime}(0)=\delta.$ Recall that Kwong proved in \cite{kw} that
for each $\alpha\in(0,1]$ there is a unique $\delta=\delta_{0}(\alpha)$ such
that $\phi$ is positive on $[0,d]$ and $\phi^{\prime}(d)=0.$

Now let
\[
s(x)=\frac{\partial\phi(x)}{\partial\delta}\text{ and }z(x)=\frac{\partial
a(x)}{\partial\delta}.
\]
Let $\delta=\delta_{0}(\alpha)$ and let $(\phi,a)$ be the corresponding
solution on $[0,d].$

\begin{lemma}
\label{lem5} $s^{\prime}(d)>0.$
\end{lemma}

Proof of Lemma \ref{lem5}:

The pair $(s,z)$ satisfies the system%

\begin{equation}%
\begin{array}
[c]{l}%
\theta^{\prime\prime}=\kappa^{2}[(\phi^{2}+a^{2}-1)\theta+2a\phi\mu+2\phi
^{2}\theta]\\
\mu^{\prime\prime}=\phi^{2}\mu+2a\phi\theta
\end{array}
\label{eq15}%
\end{equation}

and
\[
s(0)=s^{\prime}(0)=0,\,\,z(0)=0,\,\,z^{\prime}(0)=1.
\]
It is clear that $z,z^{\prime},z^{\prime\prime}$ are all positive on $(0,d]$
as long as $s>0,$ since $a>0,$ $\phi>0.$ Also, $s^{\prime\prime}%
(0)=s^{\prime\prime\prime}(0)=0,$ while $s^{\prime\prime\prime\prime}(0)>0,$
so initially, $s$ is positive. Suppose that $s(x_{0})=0$ at some first
$s_{0}>0$ in $(0,d]$, with $s>0$ on $(0,x_{0}).$ Multiply $\left(
\ref{eq15}\right)  $ by $\phi$ and $\left(  \ref{eq1}\right)  $ by $s$,
subtract and integrate. We conclude that
\begin{equation}
\phi s^{\prime}-s\phi^{\prime}|_{0}^{x_{0}}=\int_{0}^{x_{0}}(2a\phi^{2}%
z+2\phi^{3}s)dx>0\label{i1}%
\end{equation}
and applying the boundary conditions shows that $s^{\prime}(x_{0})>0,$ a
contradiction. Hence $s>0$ on $(0,d]$ and then from $\left(  \ref{i1}\right)
$ we find that $s^{\prime}(d)>0,$ as desired. This proves Lemma \ref{lem5}.

To continue our study of bifurcation from the symmetric branch, now consider
$\left(  \ref{eq1}\right)  $ with initial conditions which are possibly
asymmetric, namely
\begin{equation}
\phi(0)=\alpha,\phi^{\prime}(0)=\beta,a(0)=\gamma,a^{\prime}(0)=\delta
.\label{eq16}%
\end{equation}
Let%

\begin{align*}
p(x)  & =\frac{\partial\phi}{\partial\alpha},\,\,q(x)=\frac{\partial\phi
}{\partial\beta},\,\,r(x)=\frac{\partial\phi}{\partial\gamma},\,\,s(x)=\frac
{\partial\phi}{\partial\delta}\\
u(x)  & =\frac{\partial a}{\partial\alpha},\,\,v(x)=\frac{\partial a}%
{\partial\beta},w(x)=\frac{\partial a}{\partial\gamma},\,z(x)=\frac{\partial
a}{\partial\delta}.
\end{align*}
Then the pairs $(p,u),$ $(q,v)$ , $(r,w),$ and $(s,z)$ all satisfy the system
$\left(  \ref{eq15}\right)  .$ The initial conditions are
\begin{equation}%
\begin{array}
[c]{l}%
(p,p^{\prime},u,u^{\prime})=(1,0,0,0)\\
(q,q^{\prime},v,v^{\prime})=(0,1,0,0)\\
(r,r^{\prime},w,w^{\prime})=(0,0,1,0)\\
(s,s^{\prime},z,z^{\prime})=(0,0,0,1)
\end{array}
\label{eq16a}%
\end{equation}
at $x=0$. (Thus, these four pairs form a fundamental solution for $\left(
\ref{eq15}\right)  .$

We wish to solve the three equations
\begin{equation}
F(\alpha,\beta,\gamma,\delta)=G(\alpha,\beta,\gamma,\delta)=H(\alpha
,\beta,\gamma,\delta)=0\label{eq17}%
\end{equation}
where
\begin{equation}%
\begin{array}
[c]{l}%
F(\alpha,\beta,\gamma,\delta)=\phi^{\prime}(d)\\
G(\alpha,\beta,\gamma,\delta)=\phi^{\prime}(-d)\\
H(\alpha,\beta,\gamma,\delta)=a^{\prime}(d)-a^{\prime}(-d).
\end{array}
\label{eqFGHdef}%
\end{equation}
For each $\alpha\in(0,1],$ $(\alpha,0,0,\delta_{0}(\alpha)$ $)$ is a solution.
Starting at $\alpha=1,$ where $\delta=0,$ this solution continues uniquely as
the smooth curve of symmetric solutions as $\alpha$ decreases so long as
$J\neq0,$ where
\[
J=\det\left[
\begin{array}
[c]{lll}%
q^{\prime}(d) & r^{\prime}(d) & s^{\prime}(d)\\
q^{\prime}(-d) & r^{\prime}(-d) & s^{\prime}(-d)\\
v^{\prime}(d)-v^{\prime}(-d) & w^{\prime}(d)-w^{\prime}(-d) & z^{\prime
}(d)-z^{\prime}(-d)
\end{array}
\right]
\]
where in $\left(  \ref{eq15}\right)  ,$ $(\phi,a)$ is the solution at
$(\alpha,0,0,\delta_{0}(\alpha)).$

Since $\phi$ is even and $a$ is odd, the initial conditions $\left(
\ref{eq16a}\right)  $ imply that $p,v,w,$ and $s$ are even functions of $x,$
while $u,q,r,$ and $z$ are odd functions of $x.$ As a result, $J$ simplifies
to
\[
J=4s^{\prime}(d)(q^{\prime}(d)w^{\prime}(d)-r^{\prime}(d)v^{\prime}(d)).
\]

\begin{lemma}
\label{lemnew} For sufficiently large $\kappa d,$ $J$ changes sign between
$\alpha=0$ and $\alpha=1$.
\end{lemma}

Proof:

Lemma \ref{lem5} showed that $s^{\prime}(d)>0.$ Therefore, to obtain a
bifurcation point on the curve of symmetric solutions, we must show that
$M(\alpha)=q^{\prime}(d)w^{\prime}(d)-r^{\prime}(d)v^{\prime}(d)$ changes sign
along this curve.

For $\alpha=1$ we have $\phi\equiv1,$ $a\equiv0.$ The equations $\left(
\ref{eq15}\right)  $ and $\left(  \ref{eq16a}\right)  $ can then be solved
explicitly to show that $q^{\prime}(d)>0,$ $v^{\prime}(d)=0,$ $r^{\prime
}(d)=0,$ and $w^{\prime}(d)>0$ so that $M(1)>0.$

We now analyze $M(\alpha)$ for small $\alpha.$ First, with $(q,v)$ substituted
for $(\theta,\mu)$ and $\alpha\psi$ for $\phi$ in $\left(  \ref{eq15}\right)
,$ we multiply the equation for $q^{\prime\prime}$ by $\psi,$ $\left(
\ref{eq4}\right)  $ by $q,$ subtract, integrate, and use $\left(
\ref{eq6}\right)  $ and $\left(  \ref{eq16a}\right)  $ to obtain
\[
\psi(d)q^{\prime}(d)=1+2\kappa^{2}\alpha\int_{0}^{d}(a\psi^{2}v+\psi^{3}\alpha
q)dx.
\]
As $\alpha\rightarrow0,$ $\psi\rightarrow\psi_{0,}$ where
\begin{equation}
\psi_{0}^{\prime\prime}=\kappa^{2}(h_{0}^{2}x^{2}-1)\psi_{0},\,\,\psi
_{0}(0)=1,\psi_{0}^{\prime}(\pm d)=0\label{eq17a}%
\end{equation}
and $h_{0}$ is the unique positive number such that $\left(  \ref{eq17a}%
\right)  $ has a positive solution. (Earlier we referred to the solution
$\psi_{0}$ of $\left(  \ref{eq17a}\right)  $ as $\psi,$ but now we must
distinguish $\psi_{0}$ from $\psi=\frac{\phi}{\alpha}$ for $\alpha\neq0.$)
\ Since
\[
v^{\prime\prime}=\alpha^{2}\psi^{2}v+2a\alpha\psi q,
\]
$\left(  \ref{eq16a}\right)  $ implies that $v\rightarrow0$ on $[0,d]$ as
$\alpha\rightarrow0,$ so we consider, instead, $\frac{v}{\alpha}$, and see
that as $\alpha\rightarrow0,$%
\[
\frac{v^{\prime}(d)}{\alpha}\rightarrow\int_{0}^{d}2h_{0}x\psi_{0}q_{0}dx,
\]
where
\begin{equation}
q_{0}^{\prime\prime}=\kappa^{2}(h_{0}^{2}x^{2}-1)q_{0},\,\,\,q_{0}%
(0)=0,\,\,q_{0}^{\prime}(0)=1\label{eq17b}%
\end{equation}

We proceed in the same way with $(r,w),$ to obtain, finally, that
\begin{equation}
\lim_{\alpha\rightarrow0}\frac{M(\alpha)}{\alpha^{2}}=\frac{1}{\psi_{0}%
(d)}(\int_{0}^{d}\psi_{0}^{2}+2h_{0}x\psi_{0}R_{0}dx-4\kappa^{2}h_{0}^{2}%
\int_{0}^{d}x\psi_{0}^{2}dx\cdot\int_{0}^{d}x\psi_{0}q_{0}dx).\label{eq18}%
\end{equation}
In this expression the only term we have not defined is $R_{0},$ which is
$\lim_{\alpha\rightarrow0}\frac{r}{\alpha}$ and satisfies
\begin{equation}
R_{0}^{\prime\prime}=\kappa^{2}(h_{0}^{2}x^{2}-1)R_{0}+2\kappa^{2}h_{0}%
x\psi_{0}^{{}},\,\,R_{0}(0)=R_{0}^{\prime}(0)=0.\label{eq18a}%
\end{equation}

To prove that $J$ changes sign it is again convenient to rescale, letting
$\psi_{0}(x)=g(\kappa x),$ so that
\begin{equation}
g^{\prime\prime}=(\lambda^{2}y^{2}-1)g,\,\,g(0)=1,\,\,g^{\prime}%
(0)=0,\,\,\,g^{\prime}(D)=0\label{e1}%
\end{equation}
where $D=\kappa d$ and $\lambda=\frac{h_{0}}{\kappa}.$ It was shown in the
proof of Lemma 2 that $\lambda>1.$

Making the same change of variables in $\left(  \ref{eq18}\right)  $, we must
show that for large $D=\kappa d,$%
\begin{equation}
\frac{1}{\kappa}\int_{0}^{D}(g(y)^{2}+2\frac{h_{0}}{\kappa}yg(y)P(y))dy<4\frac
{h_{0}^{2}}{\kappa^{2}}\int_{0}^{D}yg(y)^{2}dy\cdot\int_{0}^{D}%
yg(y)Q(y)dy\label{e2}%
\end{equation}
where $R_{0}(x)=P(\kappa x),$ and $q_{0}(x)=Q(\kappa x).$ Hence, we have
\begin{equation}
Q^{\prime\prime}=(\lambda^{2}y^{2}-1)Q,\text{ \ \ }Q(0)=0,\text{
\ \ }Q^{\prime}(0)=\frac{1}{\kappa},\label{rev1}%
\end{equation}
and
\begin{equation}
P^{\prime\prime}=(\lambda^{2}y^{2}-1)P+2\lambda yg(y),\text{ \ }%
P(0)=P^{\prime}(0)=0\label{rev2}%
\end{equation}
as well as $\left(  \ref{e1}\right)  .$ Multiplying $\left(  \ref{rev1}%
\right)  $ by $g$ and $\left(  \ref{e1}\right)  $ by $Q$ and subtracting and
integrating gives
\[
\left(  \frac{Q}{g}\right)  ^{\prime}=\frac{1}{\kappa g(y)^{2}}
\]
and similarly from $\left(  \ref{rev2}\right)  $ we obtain
\[
\left(  \frac{P}{g}\right)  ^{\prime}=2\lambda\frac{1}{g(y)^{2}}\int_{0}%
^{y}sg(s)^{2}ds.
\]
\bigskip From these we find that
\begin{equation}
P(y)=g(y)\int_{0}^{y}2\frac{h_{0}\int_{0}^{x}sg(s)^{2}ds}{\kappa g(x)^{2}%
}dx,\label{e3}%
\end{equation}
and
\[
Q(y)=\frac{g(y)}{\kappa}\int_{0}^{y}\frac{1}{g(x)^{2}}dx.
\]

With these substitutions, $\left(  \ref{e2}\right)  $ becomes
\begin{equation}
\int_{0}^{D}g(y)^{2}dy<4\lambda^{2}\int_{0}^{D}yg(y)^{2}\int_{0}^{y}\frac
{1}{g(s)^{2}}\int_{s}^{D}tg(t)^{2}dt\,ds\,dy.\label{e4}%
\end{equation}

It is tempting to approach this result by studying the asymptotic behavior of
$g(y)$ as $D\rightarrow\infty.$ In fact, one can show that $g(y)\rightarrow
e^{-\frac{y^{2}}{2}}$ point-wise, and further effort can refine this result.
It turns out to be a mistake, however, to study the result of substituting
$e^{-\frac{y^{2}}{2}}$ for $g$ in $\left(  \ref{e4}\right)  $, because this
function does not satisfy the boundary conditions, and this turns out to make
the required estimates much more difficult, or indeed, impossible.

Instead, we proceed directly, and this is possible primarily because using
$\left(  \ref{e1}\right)  $ we can evaluate integrals of the form $\int
yg(y)^{2}dy.$ In particular, multiplying $\left(  \ref{e1}\right)  $ by
$g^{\prime}$ and integrating by parts, we find that
\[
\lambda^{2}\int yg(y)^{2}dy=(\lambda^{2}y^{2}-1)\frac{g(y)^{2}}{2}%
-\frac{g^{\prime}(y)^{2}}{2}.
\]

Substituting this in $\left(  \ref{e4}\right)  $ and using the boundary
conditions gives
\begin{equation}
\lambda^{2}\int_{0}^{y}\frac{1}{g(s)^{2}}\int_{s}^{D}tg(t)^{2}dtds=\frac
{\lambda^{2}D^{2}-1}{2}g(D)^{2}\int_{0}^{y}\frac{1}{g(s)^{2}}ds-\frac{1}%
{2}\frac{g^{\prime}(y)}{g(y)}.\label{e5}%
\end{equation}

Substituting this in the right side of $\left(  \ref{e4}\right)  $ gives
\begin{equation}%
\begin{array}
[c]{l}%
\lambda^{2}\int_{0}^{D}yg(y)^{2}\int_{0}^{y}\frac{1}{g(s)^{2}}\int_{s}%
^{D}tg(t)^{2}dt\,ds\,dy=\\
\frac{\lambda^{2}D^{2}-1}{2}g(D)^{2}\int_{0}^{D}yg(y)^{2}\int_{0}^{y}\frac
{1}{g(s)^{2}}ds\,\,dy+\frac{1}{4}\int_{0}^{D}g(y)^{2}dy-\frac{Dg(D)^{2}}{4}.
\end{array}
\label{e6}%
\end{equation}
Using this in $\left(  \ref{e4}\right)  $ implies that the following
inequality is sufficient for our result:
\begin{equation}
\frac{\lambda^{2}D^{2}-1}{2}g(D)^{2}\int_{0}^{D}yg(y)^{2}\int_{0}^{y}\frac
{1}{g(s)^{2}}ds\,\,dy-\frac{Dg(D)^{2}}{4}>0\label{rev3}%
\end{equation}
(for large $D).$

We need only one additional fact about $g(y);$ namely, that $1\geq g(y)\geq
e^{-\frac{y^{2}}{2}}$. We already know that $g^{\prime}\leq0$ giving the first
inequality, and for the second, differentiate $\left(  \frac{g}{e^{-\frac
{y^{2}}{2}}}\right)  $ twice and use $\ \left(  \ref{e1}\right)  $ to see that
this expression, which is $1$ at $y=0,$ then increases on $[0,D].$

Using this information it is seen that the double integral on the left of
$\left(  \ref{ref3}\right)  $ does not tend to zero with $D,$ and this proves
$\left(  \ref{rev3}\right)  $ for large $D.$ This implies $\left(
\ref{e2}\right)  $ and completes the proof of Lemma \ref{lemnew}

We also need a global bound on the solutions, for fixed $\kappa,d.$

\begin{lemma}
$\label{lembnd}$ For given $\kappa$ and $d,$ there is an $M$ such that if
$\left(  \phi,a,h\right)  $ is a solution of $\left(  \ref{eq1}\right)
-\left(  \ref{eq3}\right)  $ with $\phi>0$ on $[-d,d],$ then $|\phi
|+|\phi^{\prime}|+|a|+|a^{\prime}|\leq M$ on $[-d,d].$
\end{lemma}

Proof: Letting $\psi=\frac{\phi}{\max_{-d\leq x\leq d}\phi(x)},$ we see that
$\psi^{\prime}(\pm d)=0$ and $\psi^{\prime\prime}\geq-\kappa.$ This implies
that $\psi^{\prime}$ is bounded independent of which solution is being
considered. Further, for any solution there is an $x_{0}\in(-d,d)$ with
$a(x_{0})=0.$ If $a^{\prime}(x_{0})$ is bounded then we are done, so we can
assume that there are solutions with $a^{\prime}(x_{0})$ arbitrarily large.
Since $a^{\prime}$ is a minimum at $x_{0},$ this implies that for any $A,$ and
$\varepsilon,$ there is a $\Lambda$ such that if $a^{\prime}(x_{0})\geq
\Lambda,$ then the length of the interval in which $|a|\leq A$ is less than
$\varepsilon.$

Since $\psi^{\prime}(\pm d)=0$ we must have
\begin{equation}
\int_{-d}^{d}\left(  \psi-\psi^{3}\right)  dx=\int_{-d}^{d}\psi a^{2}%
dx.\label{d1}%
\end{equation}

Since $\max_{[-d,d]}\psi(x)=1,$ and $\psi^{\prime}\geq-\kappa,$ there most be
an $\varepsilon>0$ such that for any solution of $\left(  \ref{eq1}\right)
-\left(  \ref{eq3}\right)  ,$ $\psi\geq\frac{1}{2}$ in some interval $\Omega$
of length $\varepsilon.$ For solutions with $a^{\prime}(x_{0})$ sufficiently
large we must have $\left|  a\right|  $ large in at least half of $\Omega,$
which means that the right side of $\left(  \ref{d1}\right)  $ can be
arbitrarily large, while the left side is bounded by $2d.$ This contradiction
proves Lemma \ref{lembnd}.\newline We remark that M. K. Kwong gave a proof
using Sturmian methods \cite{Kw}.

Completion of proof of Theorem \ref{th4}:\newline 

With $M$ as in Lemma \ref{lembnd}, truncate $\left(  \ref{eq1}\right)
-\left(  \ref{eq2}\right)  $ by letting
\[
g(x)=\min(x,M^{2})
\]
and considering
\begin{equation}
\phi^{\prime\prime}=\kappa^{2}(g(a^{2}+\phi^{2})-1)\phi\label{glo1}%
\end{equation}
and
\[
a^{\prime\prime}=g(\phi)^{2}a.
\]
with boundary conditions $\left(  \ref{eq3}\right)  .$ With the notation of
$\left(  \ref{eqFGHdef}\right)  $ let $\tilde{F}=(F,G,H)=\tilde{F}%
(\alpha,\tilde{\beta}),$ where $\tilde{\beta}=(\beta,\gamma,\delta).$ We use a
truncation to ensure that whatever the initial condition at $x=0$ \ the
solutions exist up to $x=d.$

Since $J>0$ when $\alpha$ is close to $1$, (when $\tilde{F}(\alpha
,(0,0,\delta_{0}(\alpha)))=0)$, this solution $t(\alpha)=(0,0,\delta
_{0}(\alpha))$ is non-degenerate and isolated and has Brouwer degree sgn $J=1$
(for the map $\tilde{F}$ with fixed $\alpha$). Similarly$,$ if $\alpha$ is
small and positive, then $J<0$ and $t(\alpha)$ has degree -1. Hence there is a
change of degree along the branch of symmetric solutions between $\alpha$
small and $\alpha=1.$ Hence, by a slight variant of Theorem 1.16 of Rabinowitz
in \cite{rab}, there is a connected set of solutions of $\tilde{F}%
(\alpha,\tilde{\beta})=0$ branching off the symmetric solution $(\alpha
,t(\alpha))$ at a point $\alpha\in(0,1)$ where $J=0,$ and this branch will
either be unbounded in $R^{4},$ or return to (0,0,0,0), or meet the symmetric
branch again.

Note that there are no solutions with $h=0,$ since then $a$ would have to be
of constant sign and we could not have $\int_{-d}^{d}\phi^{2}a\,dx=0.$ Also,
solutions cannot bifurcate from the symmetric branch near $\alpha=0$ or
$\alpha=1,$ because $J\neq0$ there.

The solutions of interest are positive. (That is, $\phi$ is positive.)
Solutions on the bifurcating branch of asymmetric solutions start off positive
as the branch leaves the symmetric branch, from continuity. From $\left(
\ref{eq1}\right)  $ it follows that solutions can only fail to be positive
along this branch if $\phi\rightarrow0$ (uniformly on $[-d,d]$ ). Also, Lemma
\ref{lembnd} shows the solutions must remain bounded, and in fact, remain
within the truncated region where $\left(  \ref{eq1}\right)  -\left(
\ref{eq2}\right)  $ apply. If $\phi\rightarrow0,$ then $a\rightarrow h(x+c).$

It is conceivable that there are several points $\alpha$ where $J=0.$ However,
the change of degree ensures that one branch does not return to the branch of
symmetric solutions, and so, by Rabinowitz's global result, must tend to
$(0,h(x+c))$ for some $c\neq0,$ that is, to a normal solution. In fact, by the
symmetry of the problem, there will be two branches bifurcating from the same
point, one tending to the unique bifurcation point from the normal solution
with $c>0$ and the other to the symmetric reflection of this solution around
$0.$ (The uniqueness of this bifurcation point follows from Theorem
\ref{unique}.) This completes the proof of Theorem \ref{th4}

\subsection{\bigskip Conclusion:}

\ \ The initial motivation for this paper was Seydel's bifurcation diagram,
Figure 1. \ Our goal was to prove that in some parameter range the problem
could have as many as seven solutions (five essentially distinct).
\ Unfortunately we have not achieved this goal. \ There are at least two
features of Seydel's curve that seem important in obtaining such a proof. \ We
would like to determine where on the symmetric branch the bifurcation to
asymmetric solutions does exist, and we want to know the direction of
bifurcation at this point. \ These remain challenges for future work.
\ However, we have verified that for large $\kappa d$ the desired bifurcation
from the symmetric branch occurs, and furthermore, there is a curve of
asymmetric solutions going from the symmetric branch to the normal state. \ We
have also shown how earlier results on bifurcation from the normal state can
be obtained without the use of detailed asymptotics for the linear problem.

\end{document}